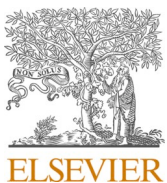
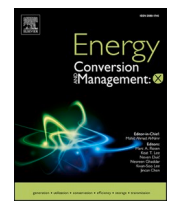

# Surrogate model of a HVAC system for PV self-consumption maximisation

Breno da Costa Paulo [a,c,*], Naiara Aginako [b], Juanjo Ugartemendia [c], Iker Landa del Barrio [a], Marco Quartulli [a], Haritza Camblong [c,d]

[a] *Vicomtech, Mikeletegi 57, 20009 Donostia-San Sebastian, Spain*
[b] *Computer Science and Artificial Intelligence Department, University of the Basque Country (UPV/EHU), Manuel Lardizabal Ibilbidea, 1, 20018 Donostia-San Sebastian, Spain*
[c] *Department of Systems Engineering and Control, University of the Basque Country (UPV/EHU), Faculty of Engineering of Gipuzkoa, Plaza de Europa 1, 20018 Donostia-San Sebastian, Spain*
[d] *Univ. Bordeaux, ESTIA INSTITUTE OF TECHNOLOGY, F-64210 Bidart, France*

ABSTRACT

In the last few years, energy efficiency has become a challenge. Not only mitigating environmental impact but reducing energy waste can lead to financial advantages. Buildings play an important role in this: they are among the biggest consumers. So, finding manners to reduce energy consumption is a way to minimise energy waste, and a technique for that is creating Demand Response (DR) strategies. This paper proposes a novel way to decrease computational effort of simulating the behaviour of a building using surrogate models based on active learning. Before going straight to the problem of a building, which is complex and computationally costly, the paper proposes the approach of active learning to a smaller problem: with reduced simulations, regress the curve of voltage versus current of a thermo-resistor. Then, the paper implements a surrogate model of energy consumption of a building. The goal is to be able to learn the consumption pattern based on a limited number of simulations. The result given by the surrogate can be used to set the reference temperature, maximising the PV self-consumption, and reducing energy usage from the grid. Thanks to the surrogate, the total time spent to map all possible consumption scenarios is reduced around 7 times.

## 1. Introduction

The increase in use of Renewable Energy Sources (RES) aiming to promote the energy transition is essential to reduce the effects of carbon emission and mitigate the risks related to climate change [1,2]. Buildings have a major impact on those emissions, and the issue of resource efficiency for building is becoming increasingly relevant [3,4]. Data models are essential to reduce problems caused by the energy transition [5,6], and particular attention should be given to the role of open data and optimisation. Operation techniques in buildings can benefit from the existent and ongoing development in data analytics [7–9]. High efficiency buildings are technically and economically feasible nowadays, however, a high penetration of weather dependent renewable energy sources poses the problem of balancing the mismatch between inflexible production and inelastic demand [10–14]. Nowadays, with the available technologies given by artificial intelligence, there is the possibility of dealing with data related to energy transition processes with a much wider perspective on sustainability [15]. What appears to be evident is the possibility of visualising synthetically (using appropriate tools) highly complex problems, represented by multivariate data structures, thereby, contributing to better decision-making processes [16,17]. Historically, in order to improve energy efficiency, buildings modelling has been developed using models based on transport phenomena. Two broad categories can be found: macroscopic and microscopic modelling [18]. These models together with parametric sensitivity analysis techniques, allow to find its optimal performance [19]. However, this methodology is highly simulation-time consuming. Instead, the use of optimisation algorithms is much more efficient regarding simulation-time [20]. Surrogate models [21–23], i.e. models that can be used to replace computationally-costly software or system, can have an important impact when dealing with optimisation problems [24,25]. There are several examples of multi-variate regression models to support design optimisation, also considering topics such as cost-optimal analysis and energy performance contracting [26].

* Corresponding author at: Vicomtech, Mikeletegi 57, 20009 Donostia-San Sebastian, Spain.
*E-mail addresses:* bdacosta@vicomtech.org (B. da Costa Paulo), naiara.aginako@ehu.eus (N. Aginako), juanjo.ugartemendia@ehu.eus (J. Ugartemendia), ilanda@vicomtech.org (I. Landa del Barrio), mquartulli@vicomtech.org (M. Quartulli), aritza.camblong@ehu.eus (H. Camblong).

Surrogate models can reduce significantly computational cost on buildings design. Future researches should be aimed at improving the simulation-time efficiency of surrogate models [20]. Special emphasis should be placed on sustainable buildings design where renewable energy sources are integrated in buildings and optimisation analysis becomes more complex [27].

This article proposes a novel way to optimise the energy supply of several devices of a university campus building, equipped with its own Photovoltaic (PV) panels, based on a surrogate model. This involves, for example, selecting when and which temperature each room has to be for a good balance between solar energy self-consumption and thermal comfort or to determine when to charge an electrical vehicle in the parking lot, choosing cost-efficient hours.

To determine the total energy consumption of a building, computationally expensive simulations are required [28]. So the main objective of this paper is to show how surrogate models can drastically reduce the simulation time needed to optimise the solar-energy self-consumption rate of a building under given conditions (internal and external temperatures, available utilities, etc.). The problem dealt with in this paper is defined in Section 2, presenting its nuances and briefly defining what is DR. Then, in Section 3, the approach taken by the proposed solution and how artificial intelligence, especially active learning, can help to reduce the consumption during peak hours are resumed. In Section 4, a proof of concept is presented, showing the problem on a smaller scale in order to experiment tools and techniques, observing their particularities. Then, in Section 5, the surrogate is used to deal with the complete problem described in Section 2. The solution is described step by step until a model that can replace the simulator is achieved. In Section 6, the results are presented, showing how much time is required with the simulator when using the surrogate and how much energy could be saved applying the DR techniques, while in Section 7 the conclusions found after implementing the surrogate are highlighted.

## 2. Problem definition

In a building equipped with photovoltaic panels, there are two main sources of energy: the PV panels themselves and the main grid. In order to consume as much PV energy as possible, the Energy Management System (EMS) can actuate over the Heating Ventilation and Air Conditioning (HVAC) system. The process of influencing the energy consumption of any system is called DR, which normally is a short-term technique and part of a long-term paradigm of energy efficiency improvement, the Demand Side Response (DSM) [29–31]. It mainly consists of actuating over electric loads, for example by means of room temperature set-points. As the conditions that affect a building energy consumption present a stochastic behaviour, due to weather, number of people inside the building, equipment usage and so on, the DR needs a control model that takes into account different operating conditions, such as winter and summer weather conditions. In a building, there is a large number of variables and parameters that influence the energy consumption. From the EMS perspective, it would be very difficult to consider all the possible combinations of variables that lead to the necessary consumption, unless a supercomputer is used. However, not all of these combinations are, actually, useful for a real situation. The problem is then to determine which combinations are useful for a real situation, and between those, which ones promote the most efficient consumption. The main challenge of the presented surrogate is to interpolate the relationship between temperature set-points and building consumption for different weather conditions.

## 3. Proposed solution

A model that can determine the consumption of an entire building is complex, regardless of the software used to create it. This paper presents the concept of active learning [32] to train a surrogate model of a simulated system. This method of machine learning improves its behaviour and its precision by adding data in the same inputs during the process of regression, instead of having a fixed number of inputs. The model is improved, in the loop, in areas where its uncertainty is high, so the regression is optimised. The diagram of Fig. 1 presents the proposed solution. The main idea is to use machine learning, specifically regression techniques, to reduce the time of creating a relationship between inputs, here rooms temperature references, and a single output, the energy consumption related to the HVAC of the building. (See Fig. 2).

Based on regression, the model presented in this paper aims to determine the best variation of the room temperature set point as a function of an outdoor temperature and an energy consumption reference for the building. So the model outputs a dataset relating energy consumption to each room temperature set-point, for different external temperatures. An algorithm is used to choose the best set-point according to two criteria: thermal comfort and PV energy self-consumption. So, in short, the active learning surrogate acts in real time to reduce the difference between PV production and energy consumption. In the end, the objective of this temperature set-point selection algorithm is to maximise the self-consumed PV energy. The main proposition of self-consumption concept is to make the energy produced and consumed locally, which means, among other advantages, reducing transmission and distribution costs, emitting less greenhouse gases and reducing final energy costs of the building. Furthermore, producing and using energy in a building itself reduces costs related to the main grid, such as profit margin and taxes, that are included in the energy bill.

## 4. Proof of concept

### 4.1. Building model: OpenModelica

To build the initial simulation model, there is a need for a software that can integrate electrical and physical behaviours. For doing this, the chosen one was the open-source modelling language Modelica. The Modelica Language is a non-proprietary, object-oriented language used to model complex physical systems containing, among others, mechanical, electrical, electronic, hydraulic, thermal, control, or process-oriented sub-components. Projects can be designed using code or helped by a block-diagram-based front end. Both methods are equivalent, and they can be extended with Python, through an Application Programming Interface (API) called OMPython.

OMPython API is an open source Python based interactive session handler for Modelica scripting. It provides the modeller with components for leveraging a complete Open Modelica (OM) modelling, compilation, and simulation environment.

### 4.2. Initial problem

Before starting with a complex model of a building, a proof of concept with a simple model was developed to verify that the integration between Modelica and Python can provide an appropriate performance for the regression process. Also, this small model was used to analyse which regressor could be the most valuable for the problem of controlling the temperature of a building. Bearing this in mind, a simple model of a thermo-resistor connected to a voltage source was built using OM. The equivalent resistance is given by two parts: a fixed and characteristic resistance value and a variable value that depends on the external temperature, so the final resistance is given by the Eq. 1:

$$R_{eq} = R(1 + \alpha(T_{heatport} - T_{ref})) \quad (1)$$

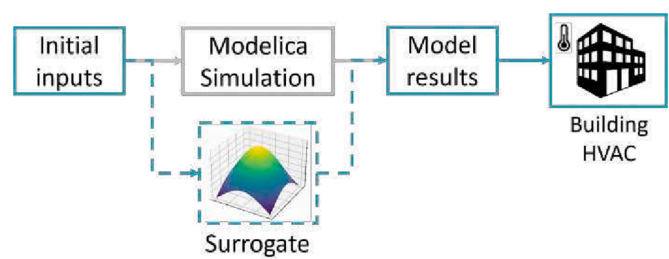

**Fig. 1.** Proposed solution. The same inputs used in OpenModelica are used in the surrogate model.

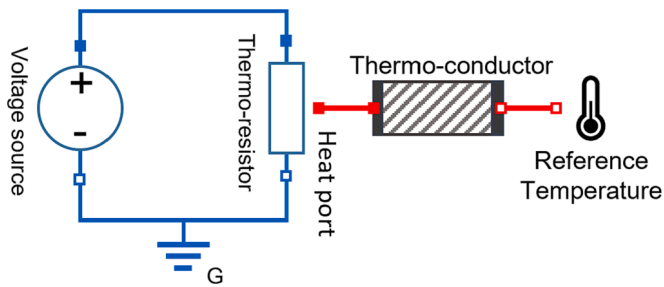

**Fig. 2.** Initial proof of concept: single thermo-resistor circuit.

Where $R$ is the intrinsic resistance, $T_{ref}$ is the reference characteristic temperature, $T_{heatport}$ is the external temperature, and $\alpha$ is the temperature coefficient of resistance, given in $K^{-1}$, which basically means that the greater the difference between external and reference temperature, the higher the equivalent resistance is.

The reference temperature is connected to the resistor through a thermo-conductor. So, the output current is dependent on the reference temperature, which has a fixed value in the initial stage, and the input voltage is a simple voltage source. The goal is to determine the function that gives the output current for an input voltage, which would be a simple function as $f(x) = ax$ if the effect of temperature did not exist. An initial voltage array is then created: 1,000 samples from 10 to 10,000 V adding a random noise. As this first part is nothing more than a proof of concept, the voltage value can be as high as 10 kV without being concerned about safety aspects or the voltage limit specification of the resistor, i.e., making this proof of concept as being a purely theoretical problem for performing a trial with the first regressor. This simple model and this amount of only 1,000 samples were chosen to make a scenario where it is possible to simulate every sample on OM to compare value by value with the surrogate result, thus giving key information about how good the regressor is. For each simulation, a value of current was collected and stored in a vector. The continuous black dots of Fig. 4 present the output current per voltage value.

### 4.3. First surrogate model

Some information can be taken from this plot. With respect to the plots in Fig. 4, as opposed to what would be expected from a single resistor circuit, the current shape of the output is not a ramp with constant slope, but it is more like an exponential curve of the type $f(x) = ae^{-bx}$. This is due to the fact that the energy coming from the voltage source is dissipated uniquely in the resistance. An increase in the voltage source means more energy being supplied to the resistor, implying more heat on it and higher equivalent resistance. So, an increase in voltage has direct-positive and indirect-negative effect on the output current. This second one becomes more important for high values, which makes the output current almost constant.

The way a surrogate model is created agrees with the diagram in Fig. 3. Briefly, a few simulations are done in OM (red stars in Fig. 4a), then, with these simulations, a first prediction is made, and its uncertainty is evaluated (blue curve in Fig. 4a). The worst point, i.e., highest standard deviation point is chosen and a new simulation is added to the training vector, which is used to perform the new prediction. When this prediction is good enough, the model is created, and it can be used instead the old one.

Standardisation of datasets is a common requirement for many machine learning estimators. They might behave badly if individual features do not more or less look like standard normally distributed data, i.e. Gaussian with zero mean and unit variance. The regression is performed in two steps. The initial one, where a set of initial input and output data are given to the surrogate to estimate the pattern between input and output. The second step is iterative, meaning that at each iteration, a new input and output pair is given to the surrogate until a threshold of minimum standard deviation is achieved. Initially, four random points were given to the surrogate. This small amount was chosen so the regressor would not fit the curve with the initial simulation, requiring more points and more simulations to achieve the minimum of uncertainty. The plot of the initial prediction is shown in Fig. 4a.

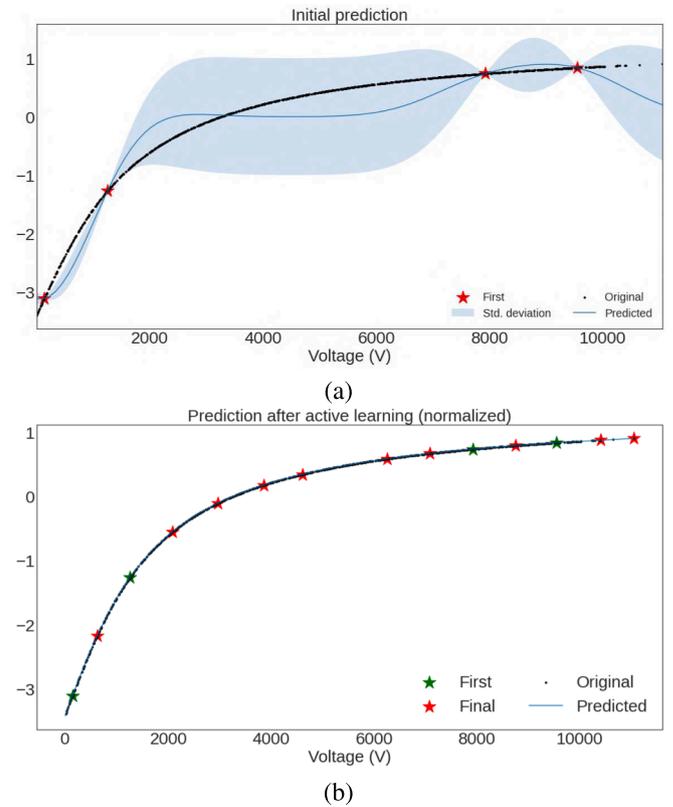

**Fig. 4.** Output current versus input voltage: in simulated and first modelled results (standardised), uncertainty decrease with increasing number of available data points.

The first prediction, as one might expect, is not accurate at all, and, for a large part of the plot, the standard deviation is fixed at its maximum of unitary value, which means that the model simply cannot determine whether the result it has found in this area is precise or not, implying a lot of uncertainty. The standard deviation is only small near the four points where the simulation was done, because, obviously, there is a simulation result on each one of these points. With this initial and inaccurate prediction, the second part of the creation of the surrogate can be started. The logic behind the iterator is quite simple. On every iteration, the point with the maximum standard deviation is chosen and added to the $X_{train}$ vector and a simulation is launched with that value of voltage. Then, the result is added to the $Y_{train}$ and the prediction is re-done. This new prediction have a different shape and the standard deviation of each point is modified. This process continues until the maximum standard deviation falls below a human-chosen threshold, which in this case was limited to 0.01. This value was chosen because for smaller limits the results do not get significantly better, but the amount of iterations required does increase a lot.

The plot in Fig. 4b presents the result of the regression after adding a few more simulations. The initial prediction points can be seen (in green). Then, ten more points are required to make the standard deviation smaller than 0.01 (shown with red stars).

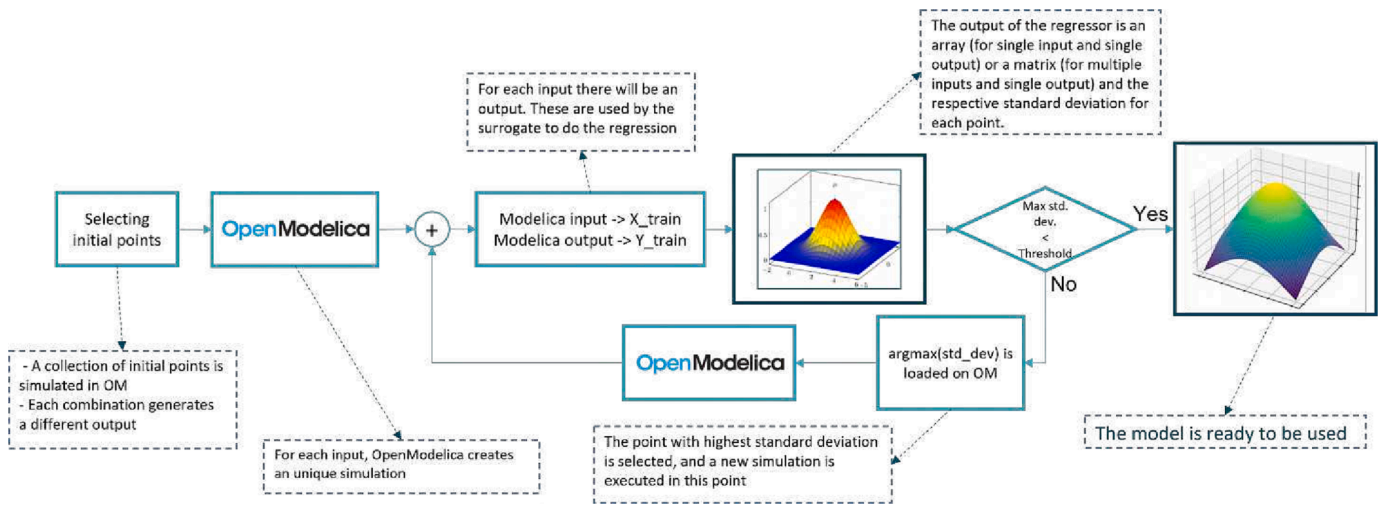

**Fig. 3.** Model creation diagram. A few initial simulations are done in OpenModelica and results are then used to create the first surrogate model. These results are improved launching new simulations until the maximum standard deviation becomes smaller than a threshold.

### 4.4. Second surrogate model: adding variable external temperature

To add more complexity to the model, the second part of it consists of adding a variable external temperature over the time. Now the resistance changes when energy is supplied by the voltage source, heating the resistance, as the previous case, but varying its final resistance value influenced by the environment temperature.

Each input vector, temperature and voltage, has a length of 50 evenly-spaced values, which means that there are $50x50 = 2,500$ different combinations. It is still possible to simulate every case to compare the surrogate performance with the simulation results. As with the single-input model, each pair of inputs is used to carry out a simulation in Modelica. The result is a 3D plot meaning a single current and a single resistance value for each pair of inputs. After simulating each condition, the result obtained is shown in Fig. 5. It can be observed that the resistance always grows in the direction of higher temperatures and higher voltages, and that the resistance, for every case, tends to stabilise when the voltage becomes very high.

The first part of the surrogate creation was launched. From the 2,500 possible points, only 10 were chosen randomly, as for the first surrogate model. This number of points was insufficient for the surrogate to achieve a good fit only with one simulation. These values were scaled and used as inputs for the regressor, which, with radial basis function kernel, made the first prediction for the surrogate. Fig. 6a shows the position of the random points in the simulated plot.

For positions that are far from the initial points, a large uncertainty is expected, and this can be confirmed with the results after the first iteration, where, especially in the edges, the uncertainty is high. In Fig. 6b, this behaviour can be seen in points such as $T_{ext} = 0\,°C$ and $V_{input} = 0$ V (left-bottom corner of the 3D plot) where the difference between simulation and prediction is clearly high. In the second part, the model is improved.

Each of these iterations consists of 4 steps. Firstly, the highest standard deviation point is collected and a simulation is released in this point. Secondly, the result is collected and appended to the $Y_{train}$ vector and this vector is re-scaled. Thirdly, a regression method fits $X_{train}$ to $Y_{train}$. Fourthly, the model is created, obtaining the matrix of 2,500 elements with the prediction of each current value.

This model can be used instead of simulating every combination. Fig. 6c shows the shape of the result with the 74 simulation points: the 10 initial points plus the added 64 points. The plot on the left is the scaled shape generated by the regressor, and the plot on the right side is the same plot after reversing the scale with values of temperature, voltage and current in proper units: °C, V and A, respectively.

As a matter of comparison, Fig. 7 presents both, simulation results and prediction results. For this amount of maximum standard deviation limit, both shapes become almost the same.

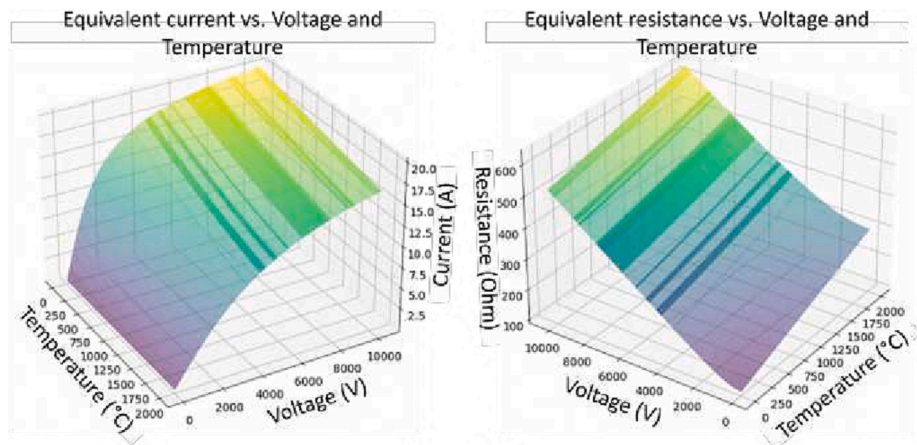

**Fig. 5.** Modelica results for the initial proof of concept. (a): Current and resistance results from software simulation. (b): Initial chosen points over Modelica simulation plot.

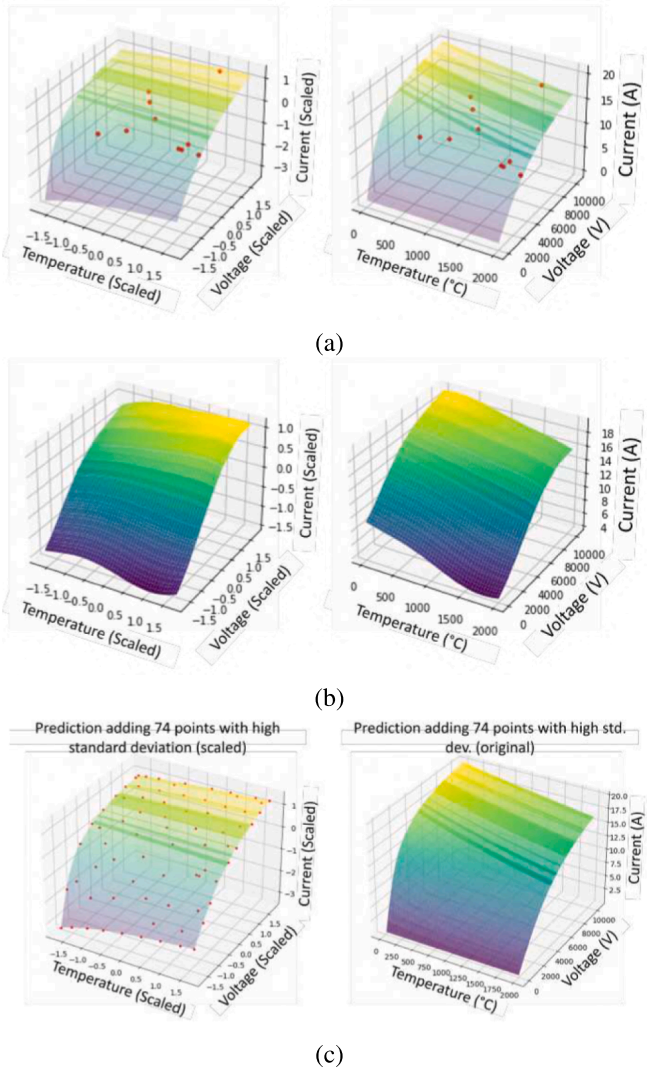

**Fig. 6.** Initial 10 randomly-chosen points over Modelica simulation plot (a), First prediction based on the initial randomly-chosen points (b), Last prediction after adding 64 more simulation points based on each standard deviation (c).

## 5. Demand response in a building

### 5.1. Problem description

This section describes a more complex and more challenging problem: replicating the HVAC system performance of a full building, considering the thermal and the electrical behaviour of several rooms. This room thermal modelling is based on a previously developed simulation model [33]. Again, using OM, each room was modelled with different thermal characteristics, in order to test if the model was able to consider these different situations. Basically, in a simulation, the variables and parameters which have influence in the temperature of a room are:

- **The starting temperature,** which is the last measured value of temperature in the room. This value will be used to decide to whether to turn the HVAC on or not [$K$];
- **The thermal conductance,** a constant which is given by the isolation of each room [$W/(m \cdot K)$];
- **The room size.** Each room is modelled as a simple volume containing air inside. This approximation is valid because a real room is indeed a volume with airflow coming in and coming out [$m^3$];
- **The mass flow rate** is the amount of air that can enter or leave the volume every second [$kg/s$];
- **The heat capacitance.** It is defined as being the amount of heat that must be supplied to the air volume to produce a change in its temperature. The unit for this block is given in Joules per Kelvin [$J/K$];
- **The HVAC heat,** the variable indirectly controlled by the DR. It supplies heat to each room according to the temperature reference, which is chosen based on simulations results. The added or removed heat is given in Joules [$J$].

Each room represents a different load that interacts with the environment in a different way from one another. Furthermore, since it is not provided information about thermal capacity of each room, it is generated heuristic data from these loads, so it is possible to recreate the interaction with the environment. The consumption of these loads depends on the external conditions and they are continually switched on and off in order to maintain the medium's temperature into a certain range.

Assuming that the modelled system is a closed system, meaning that the system does not exchange matter with its boundaries, only energy, the internal energy of the system can be determined using the First Law of Thermodynamics (Eq. 2). Thus, the internal energy of the medium (U), air in this case, changes equally to the balance of heat (Q) and work (W). Moreover, the heating processes described here are considered isochoric from a thermodynamic point of view, since the volume of the medium remains constant during the process.

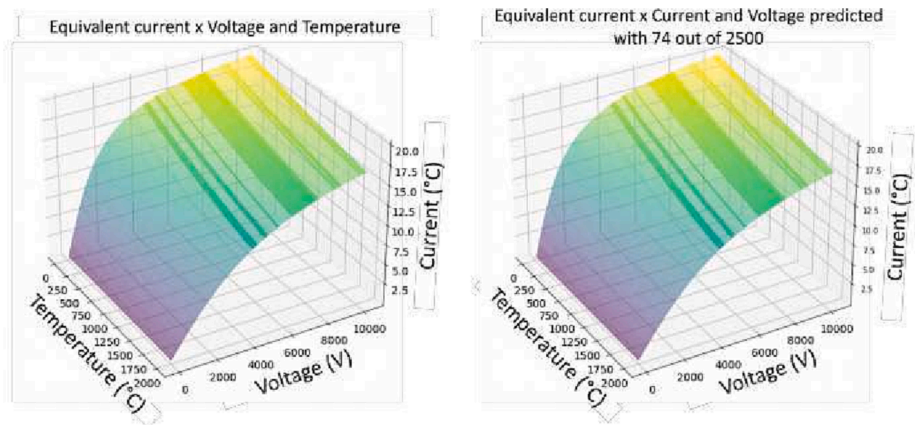

**Fig. 7.** Comparison between simulation and prediction.

$$\Delta U = Q - W \quad (2)$$

The heat flow is composed of heat inputs and outputs. In this case the input is the contribution of the heaters to the medium, while the output is the heat losses with the external air. Since the air does not change its phase through the heating process, a sensible heating process occurs. The sensible heating expression for the aforementioned isochoric heating process is described in the Eq. 3.

$$\Delta Q = m \int_{T_1}^{T_2} c_v dT \quad (3)$$

where:

- $\Delta Q$: balance of heat flows [$J$];
- $m$: mass of the medium [$kg$];
- $T_1$ and $T_2$: start/end temperatures of the medium [$K$];
- $c_v$: isochoric specific heat capacity [$J/kgK$].

The sensible heating process expression defines the balance of heat flows, this means considering the thermal losses of the medium with its boundaries, separated by a thermal insulator, and balancing them with the heat transfer of the heaters. As it is possible to deduce from Eq. 3, a positive balance will cause an increase on the temperatures, and vice versa. Therefore, any heating system must be able to ensure, at any time, the capacity to keep a positive balance between the heater's heat flow and the thermal losses' heat flow, in order to have the capacity of increasing the temperature. The thermal losses can be determined by the convective heat transfer expression, as follows in the Eq. 4:

$$Q_{Losses} = h_c A (T_{In} - T_{Out}) \quad (4)$$

where:

- $Q_{Losses}$: heat losses with the external air [$J$];
- $h_c$: heat transfer coefficient with the boundaries [$J/m^2 K$];
- $A$: Contact area of the system with the boundaries [$m^2$];
- $T_{in}$ and $T_{out}$: start/end temperatures of the medium [$K$].

### 5.2. Initial DR model

A smaller regression problem is proposed before trying to solve a complex problem such as a building formed by several rooms. The problem consists of a single-room building, which has an external temperature (environment) and a temperature reference. This initial model is a simplified version, easier to regress and faster to solve bugs during its implementation. As the number of different possible combinations are reduced, it is possible to implement a monolithic model for all cases and, moreover, simulate every combination on OM to compare the results given by the surrogate with the results given by the simulator. For the preliminary prediction, only four points were given to the predictor. They were chosen in a manner to delimit the borders of the response shape. This shape is a 3D plot, doing the relation between the two inputs, the external and reference temperatures, and the output, the hourly consumption to keep the room temperature near the reference. So the chosen values are the combinations of maximums and minimums of both arrays, according to Table 1.

The reason for choosing values to delimit the borders of the shape is to make the process of regression faster, reducing the number of future iterations. The initial surrogate model is created based on the results given by those four simulations.

In Fig. 8a, the left plot is the prediction, with a single value of consumption per each pair of $T_{ext}$ and $T_{ref}$, while the right plot is the standard deviation present in each point of plot. Apart from the four corners, where the standard deviation is zero, because the simulation result is known, the standard deviation is maximum in every other part of the plot. This means that the surrogate is completely unsure about how well the prediction is made. It simply does not know whether the result found is right or not in those points. Looking in the left plot, it can be assumed that the initial prediction is not accurate at all, because the consumption is the same, regardless of the input values.

To solve this problem, the acquisition strategy used is always to take the most uncertain point, which will be the point with highest standard deviation. Looking at Fig. 8a, this point is marked with a red dot right in the centre of the plot, with a maximum standard deviation of one. A simulation is then launched with the combinations of the temperatures that corresponds with this place in the plot.

After launching the first iteration, the uncertainty reduces considerably, from its maximum possible value, 1.0, to 0.16, which is an impressive improvement for a single added point. The new prediction is shown in Fig. 8b, and now the consumption plot has a more appropriate "appearance" with a unique value for each pair of temperatures.

In the second simulation, left plot of Fig. 8b, it is possible to observe that the consumption curve has now a better shape, having a single value for each input. But the uncertainty is still considerably high for areas that are far from the 5 simulated points. The highest one is considered again and the process restarts, launching a simulation and the next prediction with the combination indicated by the red dot in the right plot of Fig. 8b.

The matrix given by the result of the left plot of Fig. 8c is the key information that can be used instead of simulating every point on OM to decide about what reference temperature should be used.

### 5.3. Complete DR model, first approach: monolithic surrogate model

Now a complete model, consisting of 6 rooms in an university building is considered. The same regression technique is used, but with more variables, and more complexity. Instead of a 3D shape, there is a 6D or 7D shape, depending on the approach, and the number of simulations are significant, requiring the addition of a different method to optimise the computational effort, with a committee of models instead of a single model. The first six inputs represent the temperature reference for each room and a $7^{th}$ input represents the external temperature. The output is the total hourly consumption.

A modelling technique consists of creating a single and monolithic model. The results found in this surrogate should respond to every existent situation. The initial model has then 7 inputs and a single output.

The rooms temperature reference are updated hourly. Four possible reference variations are considered: the reduction of the temperature reference by 1 °C, the conservation of the precedent reference, the increase by 1 °C and the increase by 2 °C. Regarding the external temperature, in this study it covers an interval from 5 °C to 15 °C, with a 1 °C step, i.e. 11 different values. Those 7 inputs will then represent $4^6 \cdot 11$, i. e. 45.056 combinations.

This model therefore implies a very high initial effort to regress a consumption curve with this amount of combinations, but, after the regression, the results can be exploited and there will be no need to create new models. However, on the other hand, if for some reason, the characteristics of the building or the environment change, the model would not be valid anymore and a new one should be created, requiring the same large initial effort.

**Table 1**
Starting points of initial simulation.

| Simulation points (#) | $T_{ref}$ (°C) | $T_{ext}$ (°C) |
|---|---|---|
| 1 | 0 | 5 |
| 2 | 0 | 15 |
| 3 | 3 | 0 |
| 4 | 3 | 15 |

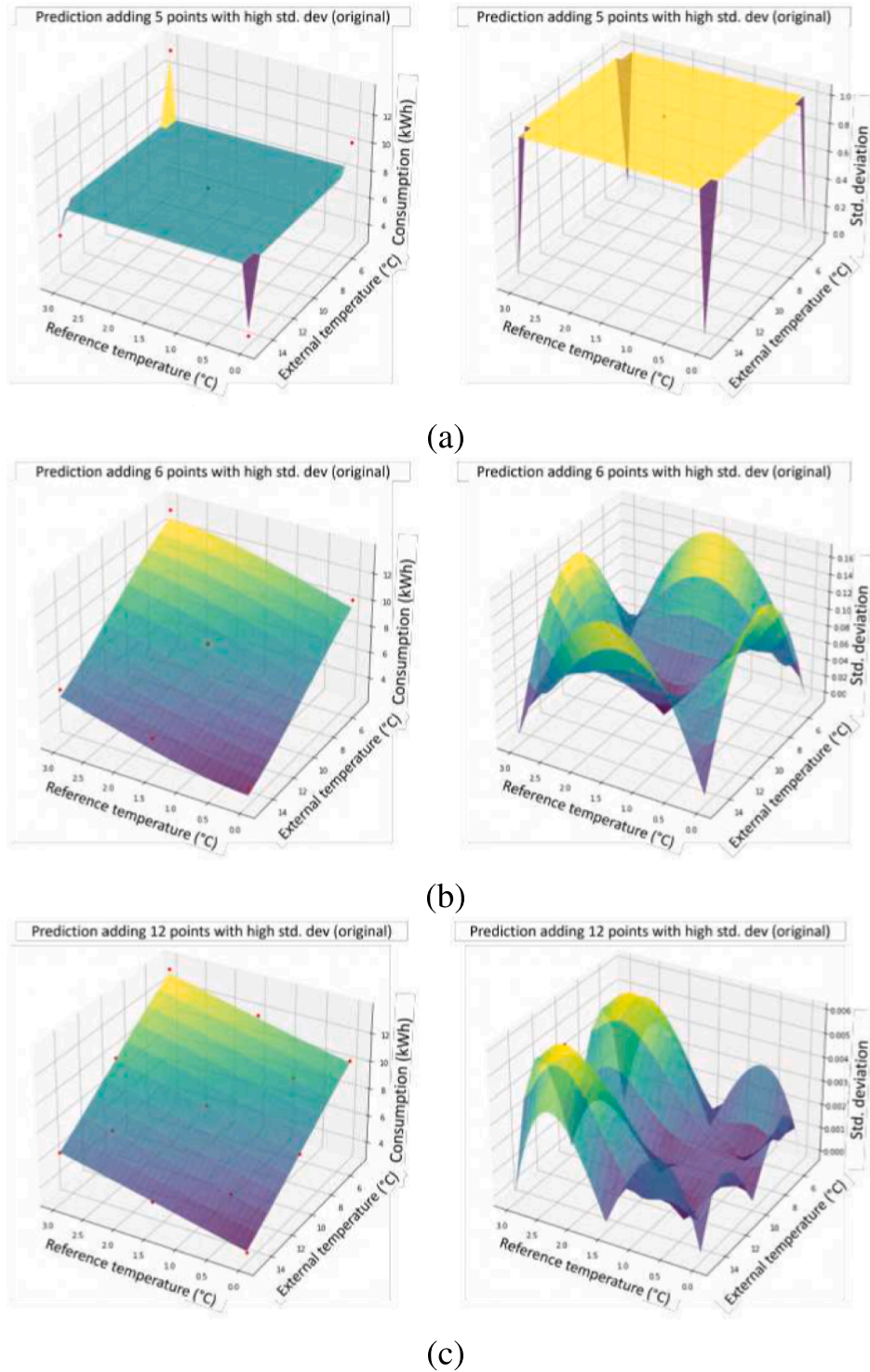

**Fig. 8.** Consumption output prediction based on external and reference temperature: Initial prediction (a), After 6 iterations (b), After 12 iterations (c).

To reduce the elevated modelling cost, choosing the initial points in a better way can lead to a substantial decrease in the number of iterations required by the active learner to achieve a reduced standard deviation. Giving an initial number of random inputs is not the best solution because it may lead to a bias in the output, in the case, for example, that all the initial input data is concentrated in a small parcel of the plot. Techniques using the design of experiments can help improving the way the initial points are chosen. One of those techniques, the deployed one, is complementary to the requirements of the regressor: escalating the input data before launching the regressor. The values are converted proportionally to have a mean equal to zero, and a max standard deviation equal to 1.

### 5.4. Complete DR model, second approach: committee of surrogate models

The second possibility is to have a committee of models. In this research study, a model for every value of external temperature with a precision of 0.5 °C is created. The same premises are taken in this strategy as before, but, instead of 7 different inputs, this strategy reduces the problem in one dimension. The external temperature is no longer an input. It is used to classify the models. Each model have then 6 inputs. The temperature reference variation is limited to 4 possibilities, as in the monolithic problem: reduction by 1 °C, conservation of the same value or increase by 1 or 2 °C. Four possibilities and 6 dimensions represent, then, $4^6 = 4096$ different combinations. So, the cost for creating a model is reduced to 10 % when compared to the monolithic model, but in this case, it is valid for a very strict interval. The main advantage of this model is the fact that there is no need to invest a large initial computational effort to have the first results. As each model is much smaller compared to the monolithic one, it can be created faster. The effort of creating new models is distributed over the time.

In the first few hours, the learning curve is expected to be steep, because the committee does not have a lot of members. Almost every hour, a different external temperature value is expected, requiring a new model. But, as the time goes on, the learning curve becomes flatter and converges to a final number. All the cases are covered.

There is another advantage. If, for any reason, the models are no longer valid anymore, the committee of models can be erased, and the learning process restarts model by model. So, compared with the monolithic model, in the worst case, the amount of lost effort will be the same, and in the mean case, it will be lower.

## 6. Results

Firstly, the results obtained for the monolithic model are analysed. The focus is on the effort of model creation rather than on how the model is used by the EMS. Then, the same analysis is done with the committee of models. It studies how the number of models increases over the time, and how the behaviour of energy consumption changed during the days when there is enough solar energy production. Finally, the obtained results about the amount of saved energy are pointed out.

### 6.1. Computational effort needed for a monolithic model

For the monolithic model, in a first simulation, all the 7 inputs are chosen fully at random. Over 45,000 different possible combinations, $2^7(128)$ are picked to be used as $X_{train}$ and fed to the regressor. As those values are random, no advantage can be taken by choosing "good" and important initial values. Not choosing the points implies a smaller computational effort in the beginning, theoretically, because no energy is needed to decide which initial values should be taken. But, this negligence also implies a more expensive iteration process in the second part of the regression, where more iterations are required to achieve the minimal threshold for standard deviation. On the other hand, when choosing the initial points based on a strategy, or using a design of experiments, the initial cost can be more expensive, because a technique must be deployed to select the points, but later, during the most expensive part of the regression (iterations), the number of times crossing the while loop decreases considerably. Fig. 9 shows the difference between random initial points and selecting points using the design of experiments for the evolution of standard deviation over the increase in the number of added simulations.

For both cases, the graph starts at 128 because this is the number of initial inputs. For the blue line, representing random initial numbers, more than 137 points are needed, apart from the initial points, meaning a process consisting of updating the parameters in Modelica, simulating and collecting the results, scaling and fitting the $X_{train}$, and creating a prediction repeated 137 times. Comparing the total number of simulations with the total number of possible combinations (265 against 45,056), the result is quite good and promising, but this value can be improved, as shown by the green curve. The regression represented by the green curve is obtained selecting the borders of the shape. It is not possible to represent it graphically, but a 7D matrix works in the same way as a 2D or 3D matrix, but simply having more dimensions. For that matrix, a combination of the minimums and maximums of each input is made, leading to 128 initial inputs: $2^n$, n being the number of dimensions. The number of iterations required for this strategy is only 17, reducing by more than 87% the number of times crossing the while loop. For this strategy, 145 simulations are done to achieve a standard deviation smaller than 0.01. This represents around 0.32% the total number of combinations. But this model, even though it is supposed to be a generic one, valid for all conditions, is useful only for temperatures between 5 °C and 15 °C. For a model valid for all possible temperatures, for example between 0 and 35 °C, the number of simulations required is extremely high to regress (between 500,000 and 600,000 different combinations), so it has been decided to leave the monolithic model just as a proof of concept and to move straight on the committee of models, which seems more promising.

### 6.2. Computational effort needed for a committee of models

Using the committee of models approach, differently from the monolithic model, this high number of simulations can be achieved because the simulations are distributed over the time. For the presented problem, the observed time interval is 15 days, or 360 h. Fig. 10 shows the behaviour of the forecasted temperature for this period, collecting the temperature profile directly from the meteorological agency Euskalmet.

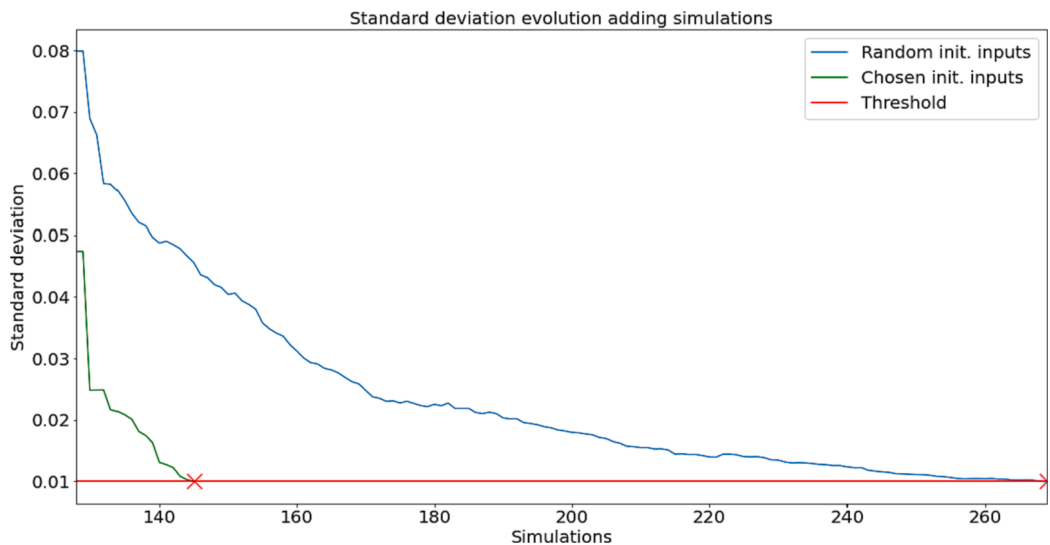

**Fig. 9.** Iterations required to achieve standard deviation threshold.

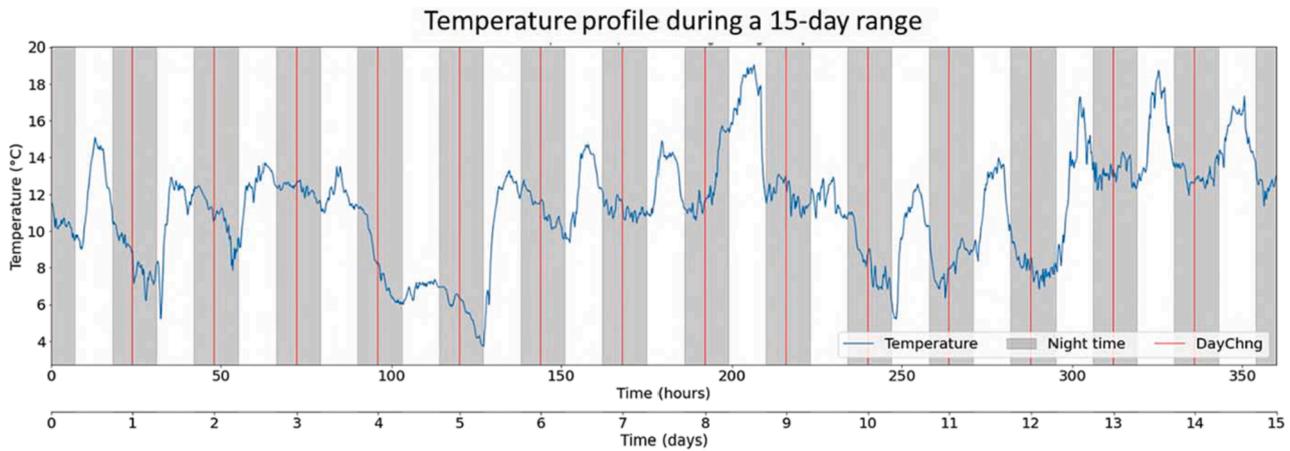

**Fig. 10.** Temperature profile during a range of 15 days.

Observing temperatures, it is noticeable that they tend to be concentrated between 9 °C and 14 °C. So the models learn mostly from this interval at the beginning. But, some peaks of higher temperatures and valleys of lower temperatures also occur, requiring an adaption in the committee of models. So, after a first and sharp learning curve, a few more models are added to the committee during the following days, especially in day 5 and day 8, where the peaks and valleys are more evident for the first time.

The number of simulations also grows at a similar rate as the number of models, requiring around 250 simulations to build each model. So, instead of simulating many times at the beginning, to later exploit the results, the first results can be exploited with just 250 simulations. Fig. 11 shows the evolution in number of models in the committee and the number of simulations to create those models.

For the total number of simulations during the 15-day period, around 7300, most part of them are carried out in the first two days (5000 simulations). This fact means that in the beginning the committee learns a lot to then use the results. Also, around 66 % of the models are created in the first 2 days, 20 models out of 30. Then, for the next 3 days, the committee does not add a new member, because the conditions of those three days are known by the existing members. This happens again after adding 3 new models in the 5$^{th}$ day, and after the 8$^{th}$ day, no new models are added and for almost half of the whole period, no effort is done to create models. The effort is concentrated in exploiting the results.

The effort of the system is divided in two parts: exploration and exploitation. Exploration is the effort used to learn and to create models, while exploitation is to use to make decisions about how to control the temperature to maximise self-consumed PV energy. Obviously, the effort in exploration is necessary, but if it is possible to carry out the exploration part with the least effort possible and to concentrate this effort in exploiting the results, better decisions will be made. But this is a trade-off, if the model is created saving computation effort in the regression time, the model will not be accurate and so the results given by the surrogate will not be compatible with the reality of the problem. But, on the other hand, if the computation to create the models is large, each model, and so the committee, will be too complicated to make the model viable. The idea is then to create a model by paying a fair computational price to achieve good results.

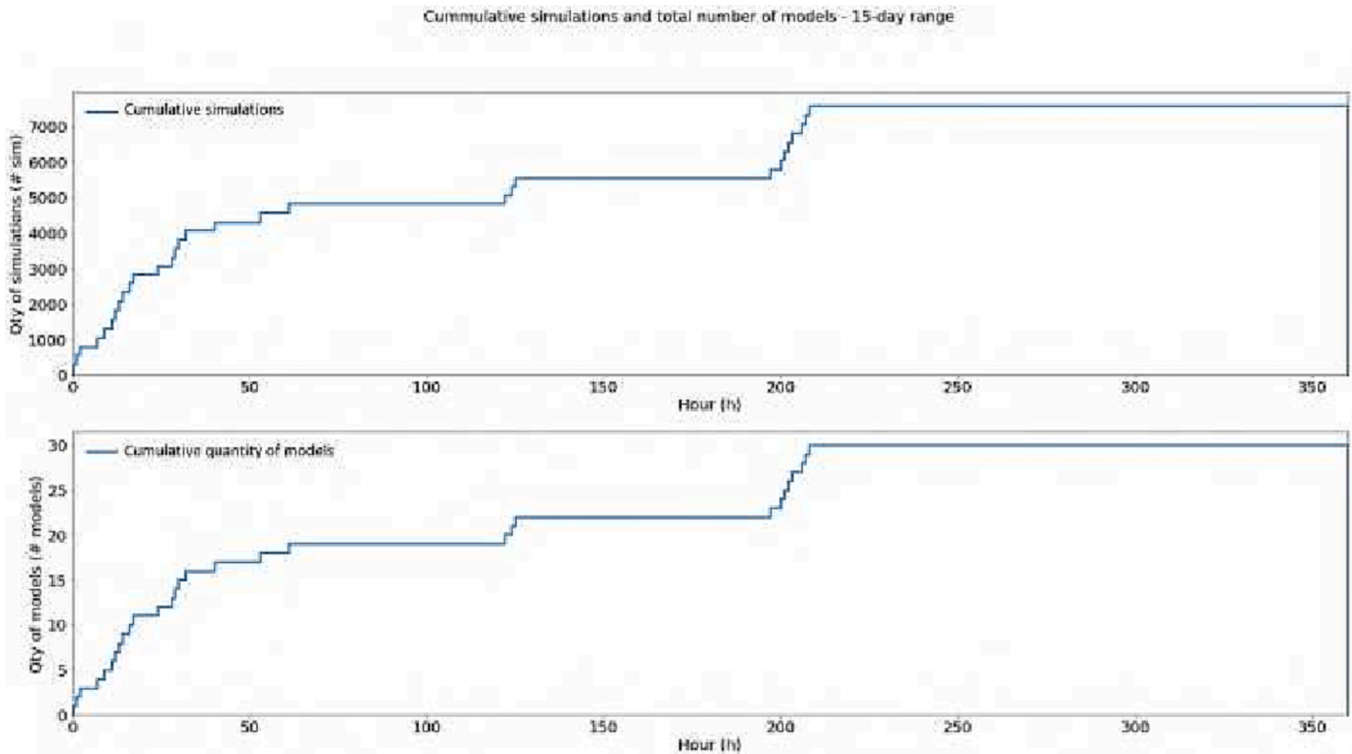

**Fig. 11.** Cumulative simulations and models.

### 6.3. Analysis of the use of committee of models to improve the self-consumption

In the exploitation part, the idea is to choose the best combination to maximise the PV energy self-consumption. For this, the use strategy selects all the possible combinations that have a consumption near to the available solar energy in that hour. Among this set of candidates, the best one is chosen by giving priority to the coldest room in the current hour. So, the set of combinations are sorted according to the coldest room, then to the second coldest room, then to the third one and so on. Doing this, all the rooms would have a similar temperature the whole time and the self-consumption is maximised. But, if for some reason, there is no production availability, the reference is set to the minimum temperature that ensure the comfort inside the building. Also, at the beginning of the day, at 6:00am, the temperature reference is increased regardless of the solar production in order to make the inner temperature comfortable before 8:00am, when people arrive in the building. Fig. 12 represents the hourly energy consumption for a period of 24 h.

It can be noticed by the orange and green curves that a peak of consumption happens in the morning. This is because the building has not been heated during the whole night. But then, when the working hours start, the building has already a temperature above 20 °C and during the interval between 8:00am and 6:00 pm, the temperature reference is chosen based on the availability of solar generation. So the surrogate tracks this energy and adjusts the temperature to maximise the self-consumption. During the first 24-h interval, the consumption of produced solar energy using the surrogate is higher than when the reference temperature is set to be a constant value during the entire day. But this cannot be seen as a final result, because specifically on this first day the surrogate behaved worse than a uncontrolled HVAC. On most of the other days the energy saved was considerably higher when the control was implemented. The comparison for the first 24-h interval is presented:

- **Consumption using surrogate:** 35.75 kWh
- **Consumption without any control:** 34.64 kWh

One can think that the surrogate increases the consumption, but in fact this is only for a short period of time, where unpredictable conditions can appear. For a longer period, such as the 15-day period, it is easier to notice the difference. Fig. 13 shows the consumption curve for this period.

The peak of consumption happens every day and this cannot be avoided, since, for a period of 12 h without any heat, the building loses a fairly amount of energy to the external environment. So, every morning the building must be heated to have a temperature around 21 °C at 8:00am, when the first people arrive. However, during the day, it can be seen in the plot that the orange curve tracks the blue curve always when possible, representing a local consumption, while the green curve simply does not have a pattern. When there is not enough solar energy, such as for example days 10 and 11, the surrogate cannot do anything but set the temperature at a fixed value and the consumption on those days is fed from the main grid. The comparison of the total consumed energy in that period is presented:

- **Consumption using surrogate:** 494.54 kWh
- **Consumption without any control:** 565.38 kWh

For this period, the reduction in total consumption is considerable, 12,5 %. But more than energy reduction, the main importance is the fact that most of this energy is produced and consumed locally, and the grid is used only as a last option.

Another enhancement achieved was to improve the Self-Consumption Rate (SCR) and the Self-Sufficiency Rate (SSR), which are indicators that represent, respectively, the amount of the produced energy that is currently being consumed locally, and the amount of energy required by the local load that can be supplied by the local generation. Both indicators are, normally, represented in percentage. The SCR is given by the Eq. 5:

$$SCR = \frac{\sum(E_{produced} - E_{exported})}{\sum E_{produced}} \quad (5)$$

where $E_{produced}$ is the energy currently being produced and $E_{exported}$ is the exceeding energy that is going to the grid.

In the same way, the SSR is given by the Eq. 6:

$$SSR = \frac{\sum(E_{produced} - E_{exported})}{\sum E_{consumed}} \quad (6)$$

where $E_{consumed}$ is the energy consumed by the building. Having a high SCR means that almost all the energy generated is being used to supply its own load, and the purpose of maximise the PV utilisation is being achieved, but it also can indicate that the energy produced is not enough for the proposed load, so all of it is being consumed and more energy from the grid is required. While the opposite can represent two different things: one is that the actual load is not high, compared to the generation, so most of the produced energy can be sent to the grid. Or, a low SCR can mean that the generation is so high at the current point that it exceeds significantly the demand and most part of it goes to the grid, instead of being used locally. About the SSR, having its percentage near 100 % means that the production is matching with the demand and less energy is required to be supplied by the grid. While having a low SSR means that the load is way heavier that the actual local energy production in a given period of time, so most part of it should be supplied by grid. As it can be seen in the Fig. 14, that compares both indicators: using the AI method and not using any kind of control (apart from setting a fixed temperature for the whole day), the reduce in the use of energy coming from the main grid is considerable. The blue filled lines represents the SCR and the SSR when the AI controller takes action to select the reference temperature, whiles the pink filled lines represent the hourly SCR and SSR when there is not any kind of control implemented. The red vertical lines represent the night time, i.e., the day change, when there is no solar generation. The night part of the plot was reduced to make it visually better.

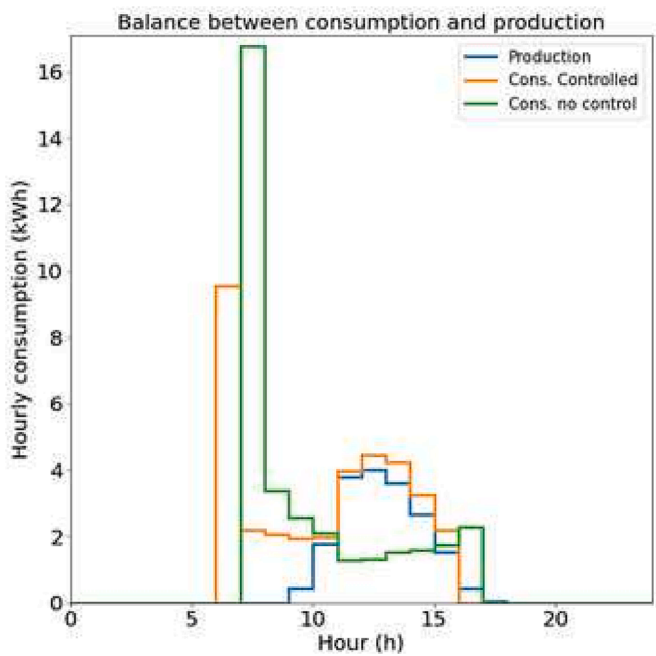

**Fig. 12.** Energy balance for a 24-h interval.

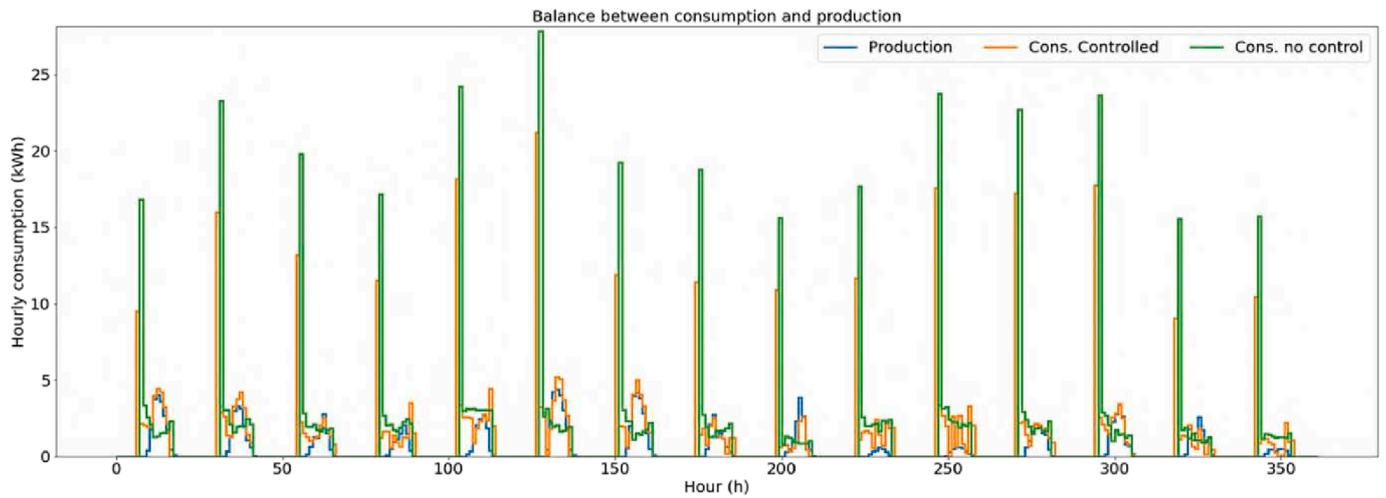

**Fig. 13.** Energy balance for a 15-day interval.

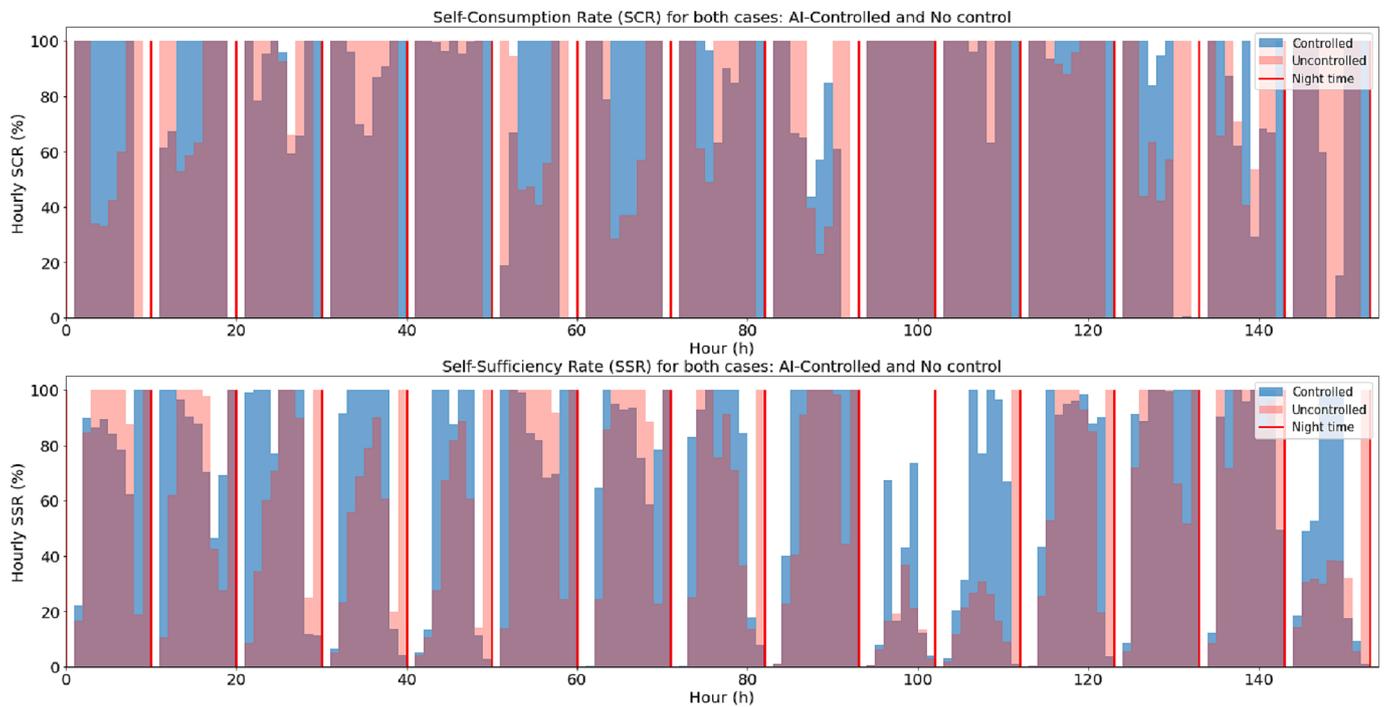

**Fig. 14.** SCR and SSR for both cases: AI-Controlled and No control.

Both indicators had a significant increase in their rates, so less energy came and left from and to the grid, which was the main goal of the developed tool. The image shows that for most of the days, the SCR increased and stayed near its maximum, meaning that the energy were not distributed, but using the heat the building. Considering the whole 15-day period, for the case of the Self-Consumption Ratio, the difference in the mean value was:

- **Mean SCR using surrogate:** 0.8349

- **Mean SCR without any control:** 0.7874

6.03 % more energy was used locally when the active learning surrogate was into action, comparing to a non-controlled HVAC system. This is a great achievement, especially when the SCR was already high without any control: it became more than 90 % in at least 10 out of the 15 days of the window time of this experiment. This means that the greatest part of the energy does not leave the building at all. About the Self-Sufficiency Rate, the improvement was considerably important, after implementing the presented control technique. The comparison between the mean SSR with and without the control is shown:

- **Mean SSR using surrogate:** 0.6309
- **Mean SSR without any control:** 0.5125

The improvement of this indicator was even more significant, 23,10 %, representing that mostly of the load is fed with local energy. It can be said that is the most important result of the experiment, especially because of the amount of improvement and the amount of saving that the surrogate can represent for the building in terms of energy cost.

As an analysis of performance, the process of regression for a 15-day period is implemented varying the threshold to verify the trade-off between regression time and accuracy. Having a bigger threshold implies a lower accuracy but, as can be seen in Fig. 15, a faster regression and a lesser computer effort. The threshold presented in the x-axis is the maximum standard deviation of the result matrix given by the regressor. Having a threshold of 0.01 consumes almost 6 more time and requires around 3.5 more simulations than having a threshold = 0.03. But this relationship changes for larger thresholds where the amount of resources is almost the same, for threshold 0.03 and 0.06, and this is because of the initial and non-iterative step of the regression process, as explained below.

According to it in Fig. 15, the amount of total simulations converges to a horizontal asymptotic line of minimum required simulations, and for 30 models describing the consumption behaviour of 30 different values of temperature, this value is equal to 1920 simulations. As described in the problem definition, the first step of the regression starts with some initial inputs, and to make the regression process faster, the chosen points delimit the borders of the input matrix. For example, if there are only two inputs, let's say room R1 and room R2, there would be an input matrix of 2 dimensions, and four inputs would be required to delimit this matrix, ($[R1_{min},R2_{min}]$, $[R1_{min},R2_{max}]$, $[R1_{max},R2_{min}]$, $[R1_{max},R2_{max}]$), so for each model, at least four simulations would be deployed. The presented building model having 6 inputs, it has a 6D matrix as input, which means $2^6 = 64$ initial inputs per model, implying $30(2^6) = 1920$ simulations for the thirty different temperatures evaluated in the time range. For non-constricting threshold, such as 0.05 and 0.06, the regression does not even have to initiate the iterative part, finishing the process using only the initial inputs. That is why the curve of total simulations converges to an asymptotic line as the threshold increases.

## 7. Conclusions

The use of surrogate and new machine learning techniques to improve DR can be a game changer in energy transition. In the energy industry, the amount of available data is a key asset. Several historical data forecasts exist, but the issue is how to process all this data and how to make decisions based on it.

In order to prove that the use of committee of surrogate models can bring many advantages to DR strategies, we have replicated the HVAC system performance of a entire building with several rooms. Obtained results support the use of committee of surrogate models in the presented scenario. As shown in Table 2, the total simulation time is reduced around 7 times, and the total number of simulations is reduced around 94 % at a cost of only 2 % error.

About this prediction error, when having a threshold of uncertainty of only 1 %, the error is minimised, representing only 2 % in the worst case, and, on average the error is less than 0.3 %. The error is calculated taking a model created by a surrogate and doing an element-wise division by the real and simulated data. If all the values were the same, the result of this division would be unitary for every element in the matrix. But the maximum value of this division was 1.02, meaning the predicted value was 2 % higher. It is a fair price to pay for the reduction in the number of simulations obtained. The presented results were made considering only a period of 15 days, for a larger period, such as one year, for example, the greater part of this time would be only exploitation rather than exploration. So, the effort done by the committee achieves a maximum, while the information about external temperature is fully covered after a certain period.

**Table 2**
Performance results.

| Parameter | Surrogate | OpenModelica |
| --- | --- | --- |
| Total number of sims | 7,500 | 122,800 |
| Simulation time | 1,831 s | 12,288 s |
| Time per simulation | 14.9 ms | 100 ms |
| Simulation rate | 67 sim/s | 10 sim/s |
| Max error | < 2 % | 0 % |

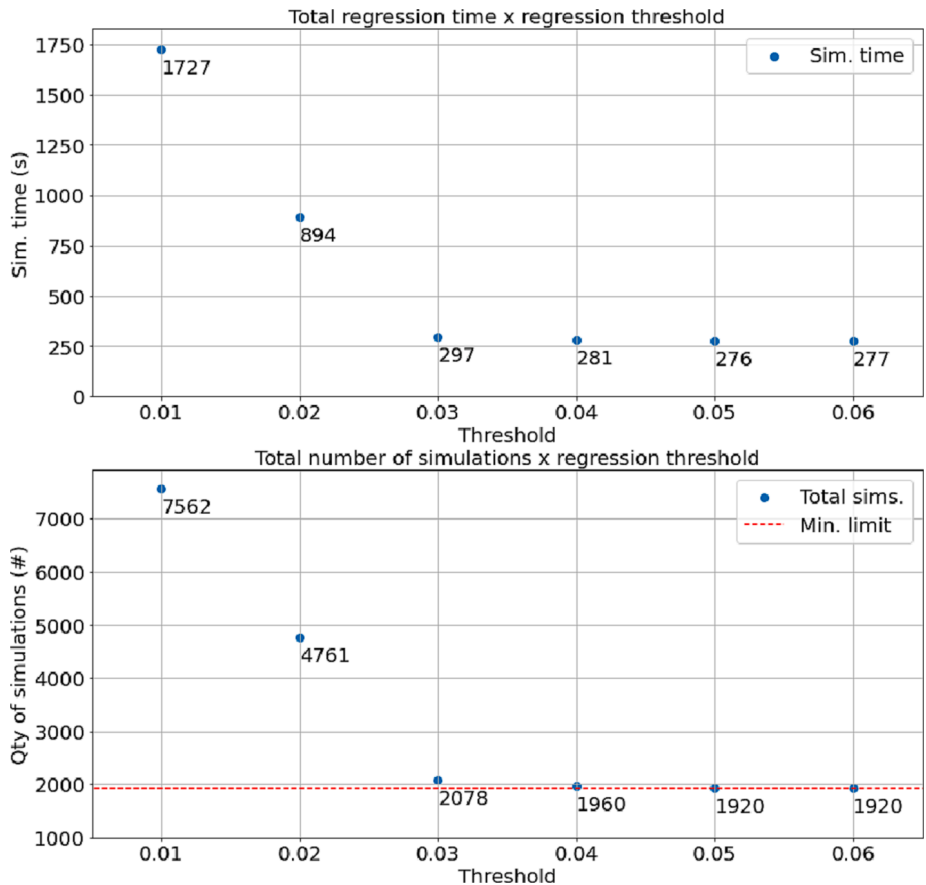

**Fig. 15.** Surrogate inference quality versus computational cost.

Although the results obtained have been analysed in depth for the chosen use case, a more generalised conclusion can be drawn from this work. The use of committee of surrogate models can also be extrapolated to other environments where large volumes of simulations are necessary to model different scenarios. Committee of models allow simulation time to be reduced at a very low cost and with very small error rates.

There are also scenarios where the implementation of these models is not appropriate. For example, in scenarios where the number of data grows exponentially or the dealing data does not have repetitive behaviour as in the case of temperature. In other words, the temperature of a building will remain within known values and therefore new models can be created to adapt to these new values. In a scenario where these values are constantly changing and do not repeat, a new surrogate model would have to be created, turning the problem into an intractable problem due to the large number of models that would have to be managed.

On the other hand, the solution presented within this document focuses on the simulations' computational costs reduction to perform an energy management strategy, assuming the simulation results as the ground truth for the surrogates. Therefore, the use of a sufficiently calibrated simulator is needed, and assumed, as a basic element to produce an acceptable performance of the whole management strategy, by reducing the propagation of error. However, this calibration is out of the scope of the present study.

**CRediT authorship contribution statement**

**Breno da Costa Paulo:** Investigation, Data curation, Writing - original draft. **Naiara Aginako:** Methodology, Formal analysis. **Juanjo Ugartemendia:** Validation, Writing - review & editing. **Iker Landa:** Resources, Writing - review & editing. **Marco Quartulli:** Conceptualization, Project administration, Funding acquisition. **Haritza Camblong:** Supervision, Writing - review & editing.

**Declaration of Competing Interest**

The authors declare that they have no known competing financial interests or personal relationships that could have appeared to influence the work reported in this paper.

**Data availability**

Data will be made available on request.

**Acknowledgements**

This work was developed as a specific task within the EKATE (EFA312/19) research project which has been supported by FEDER Interreg POCTEFA program. The project was, also, partially funded by EXPERTIA project (KK-2021/00048) and MURRIZTU (TED2021-129488B-I00).